Title: Suprathermal electrons at Saturn's bow shock

Short title: Suprathermal electrons at Saturn's bow shock

A. Masters[1], A. H. Sulaiman[2], N. Sergis[3], L. Stawarz[4], M. Fujimoto[5,6], A. J. Coates[7,8], M. K. Dougherty[1].

[1]The Blackett Laboratory, Imperial College London, Prince Consort Road, London, SW7 2AZ, UK.

[2]Department of Physics and Astronomy, University of Iowa, Iowa City, IA, USA 52242.

[3]Office of Space Research and Technology, Academy of Athens, Soranou Efesiou 4, 11527 Athens, Greece.

[4]Astronomical Observatory, Jagiellonian University, ul. Orla 171, 30-244 Krakow, Poland.

[5]Institute of Space and Astronautical Science, Japan Aerospace Exploration Agency, 3-1-1 Yoshinodai, Chuo-ku, Sagamihara, Kanagawa 252-5210, Japan.

[6]Earth-Life Science Institute, Tokyo Institute of Technology, 2-12-1 Ookayama, Meguro, Tokyo 152-8551, Japan.

[7]Mullard Space Science Laboratory, Department of Space and Climate Physics, University College London, Holmbury St. Mary, Dorking RH5 6NT, UK.

[8]The Centre for Planetary Sciences at UCL/Birkbeck, Gower Street, London WC1E 6BT, UK.

Corresponding author: A. Masters

E-mail: a.masters@imperial.ac.uk

Abstract

The leading explanation for the origin of galactic cosmic rays is particle acceleration at the shocks surrounding young supernova remnants (SNRs), although crucial aspects of the acceleration process are unclear. The similar collisionless plasma shocks frequently encountered by spacecraft in the solar wind are generally far weaker (lower Mach number) than these SNR shocks. However, the Cassini spacecraft has shown that the shock standing in the solar wind sunward of Saturn (Saturn's bow shock) can occasionally reach this high-Mach number astrophysical regime. In this regime Cassini has provided the first *in situ* evidence for electron acceleration under quasi-parallel upstream magnetic conditions. Here we present the full picture of suprathermal electrons at Saturn's bow shock revealed by Cassini. The downstream thermal electron distribution is resolved in all data taken by the low-energy electron detector (CAPS-ELS, <28 keV) during shock crossings, but the higher energy channels were at (or close to) background. The high-energy electron detector (MIMI-LEMMS, >18 keV) measured a suprathermal electron signature at 31 of 508 crossings, where typically only the lowest energy channels (<100 keV) were above background. We show that these results are consistent with theory in which the "injection" of thermal electrons into an acceleration process involves interaction with whistler waves at the shock front, and becomes possible for all upstream magnetic field orientations at high Mach numbers like those of the strong shocks around young SNRs. A future dedicated study will analyze the rare crossings with evidence for relativistic electrons (up to ~1 MeV).

1. Introduction

Collisionless shock waves are ubiquitous in space plasma environments, both in the Solar System and beyond. As with all shocks, they form wherever the speed of a flow with respect to an obstacle is faster than the speed at which information can be transferred via the medium. Flow kinetic energy is dissipated at a shock, and in the case of shocks in highly tenuous space plasmas this dissipation occurs via charged particle interactions with the electromagnetic field, rather than via particle collisions (see the review by Treumann 2009).

Key parameters that control the physics of a collisionless shock include the shock Mach numbers and the shock angle. Each Mach number is the component of the upstream flow velocity normal to the shock front (in the shock rest frame) divided by a characteristic upstream wave speed (e.g., the Alfvén speed). Shock Mach numbers (particularly the fast magnetosonic Mach number) indicate how much flow kinetic energy has to be dissipated. The shock angle, $\theta_{Bn}$, is the angle between the local normal to the shock surface and the upstream magnetic field, which strongly influences particle motion at the shock. At quasi-parallel shocks ($\theta_{Bn} < 45°$) particles can move back upstream (against the bulk flow) more easily, whereas at quasi-perpendicular shocks ($\theta_{Bn} > 45°$) upstream motion is more limited.

A major theme of research on the topic of collisionless shocks in space plasmas concerns the shock-related processes that can accelerate particles to very high energies. This is driven by the historic problem of explaining the sources of the high-energy cosmic ray charged particles that pervade space. Cosmic rays up to ~$10^{15}$ eV are thought to have been accelerated within our Galaxy, and although different theories for galactic particle acceleration to such energies have been proposed the leading model involves acceleration at the shock waves that surround young (≤1000 year-old) supernova remnants (SNRs; e.g., Blandford & Eichler 1987). This is partly because of the available energy in such systems, where a cloud of stellar debris rapidly expands and drives collisionless shocks in the surrounding plasma, and the overall supernova explosion rate.

Remote evidence that young SNR shocks are indeed capable of accelerating particles to high energies comes from radio, x-ray, and also gamma ray observations (e.g., Aharonian et al. 2004; Uchiyama et al. 2007; Reynolds 2008; Abdo et al., 2011, Helder et al. 2012). Note that although ultrarelativistic electrons represent a very small fraction of primary cosmic rays (~1% in the GeV-TeV energy range; e.g., Ackermann et al. 2010), they may dominate radiative outputs of SNRs in various (or even all) accessible electromagnetic channels. While the acceleration is thought to occur via a Fermi process (where ions and electrons bounce between converging scattering centers either side of the shock front, often referred to as Diffusive Shock Acceleration – DSA; e.g., Blandford & Eichler 1987; Drury 1983; Jones & Ellison 1991), in the absence of *in situ* measurements some crucial aspects of the acceleration process are poorly understood – in particular those related to electron "injection", and magnetic field amplification at the shock front (e.g., Bell 2013).

Interplanetary collisionless shocks are common in the continuous high-speed flow of solar wind plasma from the Sun (e.g., Russell 1985; Smith 1985), and represent an accessible natural laboratory within which spacecraft can make *in situ* observations. Spacecraft data taken during crossings of these heliospheric shocks has revealed much about the energy dissipation involved, both in heating the bulk electron and ion plasma and in accelerating a small fraction of particles to higher energies (see the review by Burgess 2007). The latter of these is most relevant for the cosmic ray source problem, despite the fact that maximum particle energies are limited by the far smaller scale of heliospheric shocks compared to their much larger young SNR shock counterparts. Until recently shock-acceleration of electrons had only been identified at quasi-perpendicular shocks (Sarris & Krimigis 1985; Gosling et al. 1989; Krimigis 1992; Shimada et al. 1999; Oka et al. 2006), and the lack of evidence for electron acceleration under quasi-parallel conditions has featured heavily in discussions of the "electron injection problem". This is the anticipated inefficiency of resonant interactions between thermal-pool electrons and magnetohydrodynamic (Alfvénic) turbulence, prohibiting any DSA at the shock front (e.g., Shimada et al. 1999). However, the implications of *in situ* results for electron acceleration at young SNR shocks have been unclear, since heliospheric shocks are considerably weaker (lower Mach number) than these far stronger examples of astrophysical shocks.

Recently reported observations made by the Cassini spacecraft during its orbital tour of Saturn have shown that the shock wave that stands in the solar wind sunward of the planet (Saturn's bow shock) is occasionally able to bridge the gap to the high-Mach number regime of young SNR shocks (Masters et al. 2011; Masters et al. 2013; Sulaiman et al 2015). This is possible because the evolution of solar wind parameters with heliospheric distance makes Saturn's bow shock one of the strongest in the Solar System (e.g., Russell 1985), and such occasions occur under rare solar wind conditions where the near-Saturn Interplanetary (solar) Magnetic Field (IMF) strength drops to ~0.1 nT and the shock Alfvén Mach number increases to order 100. On such occasions Cassini has witnessed electron acceleration at a quasi-parallel shock crossing (Masters et al. 2013), and provided evidence for shock reformation controlled by specular ion reflection (Sulaiman et al., 2015). The electron acceleration result indicates that there may not be an electron injection problem at high Mach number quasi-parallel collisionless shocks, in agreement with some theories (e.g., Amano & Hoshino 2010).

Here we reveal the full picture of suprathermal electrons at Saturn's bow shock revealed by Cassini. We show that the sum of all Cassini electron observations made during hundreds of shock crossings is consistent with electron acceleration theory that involves interactions with whistler waves excited by the reflected thermal electrons just upstream of the shock. The implication for the strong shocks surrounding young SNRs is that they may be able to inject thermal electrons into an acceleration process under any upstream magnetic field orientation.

2. Survey of suprathermal electron signatures at Saturn's bow shock

The Cassini spacecraft has been in Saturn orbit since July 2004. During the orbital tour the spacecraft has crossed Saturn's bow shock hundreds of times. The location of the shock is highly variable, and the boundary moves at speeds much greater than that of the spacecraft (Achilleos et al. 2006). As a result, multiple shock crossings are typically made on an inbound/outbound pass of any orbit where the spacecraft enters the region of space defined by the range of possible shock locations.

Two of the sensors carried by instruments mounted on the three-axis stabilized spacecraft are particularly relevant for a survey of electron acceleration at Saturn's bow shock. The first is the Electron Spectrometer (ELS) of the Cassini Plasma Spectrometer (CAPS), which detects electrons in the (lower) energy range 0.5 eV to 26 keV (Young et al. 2004). The second is the Low Energy Magnetospheric Measurements System (LEMMS) of the Magnetospheric Imaging Instrument (MIMI), which detects electrons in the (higher) energy range 18 keV to ~1 MeV (Krimigis et al., 2004). Both ELS and LEMMS have a limited field-of-view (FOV). In addition, measurements of the local magnetic field vector made by the fluxgate magnetometer of the Cassini dual-technique magnetometer (MAG; Dougherty et al. 2004) provide an essential diagnostic of shock structure.

This study is motivated by recently reported observations made by Cassini at a single (quasi-parallel) shock crossing, where suprathermal electrons were detected (Masters et al. 2013). The high-energy electron signature of this event is most pronounced in data taken by the more sensitive LEMMS sensor, with a peak intensity in all LEMMS electron energy channels that is effectively coincident with the time the spacecraft crossed the shock front. The ELS data taken during this event reveal a clear signature of the shock crossing in the thermal electron distribution. During this particular event the spacecraft was rolling, improving the FOV of both sensors and covering all pitch angles over the duration of the LEMMS signature. The lack of evidence for an associated modulation of LEMMS channel intensities suggests that the shock-accelerated electron population is sufficiently isotropic that its detectability is independent of sensor FOV, consistent with observations of electrons accelerated at Earth's bow shock (e.g., Gosling et al. 1989). Based on this result we surveyed all LEMMS electron data taken during Cassini bow shock crossings and identified cases where a signal was observed by LEMMS that is temporally correlated with the time of the shock crossing (i.e., where channel intensities change at the approximate time of the crossing, or where intensities are at a local maximum).

Magnetic field data taken during the mission to date reveal 871 unambiguous bow shock crossings (Sulaiman et al. 2015). Electron data taken by LEMMS is available for 856 of these 871 crossings. Sunlight contamination masks any shock-associated signature at 348 of these 856 events. Figure 1 shows the locations of the remaining 508 crossings. Figure 1a shows that Cassini bow shock crossings occur across the dayside shock surface, and Figure 1b shows that these crossings were predominantly made at low latitudes. The prevailing IMF orientation at Saturn orbit is approximately parallel/antiparallel to the *y*-axis, meaning that Cassini generally encounters a quasi-perpendicular shock, as previously reported (e.g., Masters et al. 2011).

A two-hour-long time series of the intensity of the lowest LEMMS electron energy channel (18-36 keV) centered on the time of each of these 508 shock crossings was analyzed. Channel background intensities are updated every few months, where a one-hour period is selected when only background was measured. The intensity of the lowest energy channel was approximately at the associated background level surrounding all 508 crossings. However, the mean value of the fluctuating background measured near each crossing can differ from the predicted level (updated on a timescale of months). To identify candidates for solar wind electron acceleration at Saturn's bow shock we required the presence of a signal temporally correlated with the shock crossing time (see above) where the peak intensity was greater than the mean intensity in an adjacent one-hour-long window plus five standard deviations. This condition was met at 31 crossings. These are shown as colored symbols in Figure 1, whereas crossings without a LEMMS electron signature are shown as gray dots. Blue symbols indicate the 28 of the 31 crossings where only the lowest LEMMS electron channels (<100 keV) were above background. Red symbols indicate the three events where all channels were above background (up to ~1 MeV).

These 31 events could be cases of shock-acceleration of solar wind electrons that is the focus of this study (discussed above), but could also be cases where electrons that had escaped from inside Saturn's magnetic field cavity (magnetosphere) were observed at the time the spacecraft crossed the bow shock by coincidence. Cassini has observed "leaked" magnetospheric ions in the near-Saturn solar wind (Sergis et al. 2013), and a clear population of leaked magnetospheric electrons was identified for the reported case of electron acceleration at Saturn's quasi-parallel bow shock (Masters et al. 2013), where it was successfully separated from the population of shock-accelerated solar wind electrons.

These two scenarios can be differentiated by inspecting the magnetic field data taken during each of the 31 shock crossings with a LEMMS signal. Leaked magnetospheric electrons are tied to magnetic field lines (the gyroradius of a 20 keV electron in the downstream solar wind is ~1 Saturn radius), and so a magnetic connection between the event location and the magnetopause boundary of Saturn's magnetosphere is necessary for leakage to be plausible. Combining semi-empirical global models of Saturn's bow shock and magnetopause (Kanani et al. 2010; Went et al. 2011) with the mean magnetic field in a five-minute window immediately downstream of each crossing indicates whether there was such a magnetic connection at the time. This is indicated in Figure 1, where the unfilled colored symbols correspond to events where there was a magnetic connection, and thus a leakage interpretation is plausible (but not conclusive). The leakage interpretation can be ruled out for 26 of the 31 events (filled colored symbols), confirming that these are examples of solar wind electron acceleration by Saturn's bow shock.

3. Observations made during example shock crossings

Data taken by MAG, LEMMS, and ELS during three example crossings of Saturn's bow shock are shown in Figure 2. Example 1, shown in Figures 2a through 2c, is the first shock crossing made by Cassini (in June 2004), and has a signature in all three data sets that is typical. The MAG data (Figure 2a) shows a relatively sharp transition from upstream (weaker magnetic field) to downstream (stronger magnetic field), which is characteristic of a quasi-perpendicular shock. This is supported by combining the time-averaged magnetic field vector over a 5-minute interval immediately before the sharp field strength increase with a local normal to the shock surface predicted by a semi-empirical model (Went et al. 2011), which is preferred to other shock normal determination methods (Horbury et al. 2002; Achilleos et al. 2006). This gives $\theta_{Bn}$ ~ 70° in the case of this crossing (Masters et al. 2011). This typical example is shown as a gray dot in Figure 1 because all LEMMS electron channels were at background surrounding the crossing time (Figure 2b, where channel backgrounds have been subtracted). Upstream of the shock front the ELS sensor measured an above-background, mixed population of ambient solar wind and spacecraft photoelectrons at energies below ~10 eV, whereas downstream the ambient population is clearly resolved at higher energies (up to ~300 eV).

Example 2, shown in Figures 2d through 2f, is also not associated with a LEMMS signal (gray dot in Figure 1). However, the magnetic structure of this shock is less typical. This is in fact an example of two shock crossings, upstream-downstream (inbound) at ~06:20 Universal Time (UT) and downstream-upstream (outbound) at ~06:55 UT on 25 October 2004. At both shock crossings the shock front is less clear than in the first example, and there is a greater level of upstream magnetic field fluctuations (Figure 2d). This is indicative of a lower shock angle, consistent with the calculated value of $\theta_{Bn}$ ~ 60°. The LEMMS and ELS signature of this pair of crossings is qualitatively similar to that of the first example.

Example 3, shown in Figures 2g through 2i, is one of the 31 crossings where a shock-associated LEMMS signature was identified (filled blue square in Figure 1). The magnetic field structure of this inbound crossing on 14 June 2007 is typically quasi-perpendicular, with $\theta_{Bn} \sim 90°$ (Figure 2g, note that the differing time period of field fluctuations between the three examples is likely caused by different speeds of the shock surface as it moves over the spacecraft). Figure 2h shows the above-background intensities in the lowest three LEMMS energy channels (up to ~100 keV) that began at the approximate time of the shock front crossing (when the magnetic field strength rapidly increased) and continued for ~8 minutes after this time (i.e., measured immediately downstream of the shock). The ELS signature is essentially typical, although the downstream thermal electron population extends up to higher energies than in the other examples (although still of order 100 eV).

Figure 3 shows a two-minute-averaged electron energy spectrum (combining ELS and LEMMS) for each example shown in Figure 2. The intervals were chosen immediately downstream of the shock front in each case. Note that no background-subtraction has been applied to either the ELS or LEMMS data in this figure, and that both the energy range and background level of each LEMMS electron channel is indicated by dotted lines, with the measured channel intensity given as data point. As indicated in Figure 2, the ELS spectrum of example 3, where a LEMMS signal was identified, produced above-background intensities up to higher energies than in Examples 1 and 2 where no associated LEMMS signal was identified.

One of the three shock crossings shown as red circles in Figure 1, which were associated with LEMMS signals where all channels were above background (up to ~1 MeV), is the previously reported example of electron acceleration at a very high-Mach number quasi-parallel shock (Masters et al. 2013). The LEMMS data for the other two cases will be discussed in a forthcoming dedicated study that compares electron acceleration efficiency at quasi-perpendicular and quasi-parallel shocks in detail.

4. Discussion

The *in situ* data analysis results presented in Sections 2 and 3 show that instrumentation carried by the Cassini spacecraft is rarely able to resolve a signature of electron acceleration at Saturn's bow shock, and even more rarely with continuity in energy form eV to MeV energies (i.e., all intensities above background; Masters et al. 2013). However, the Cassini data has provided 31 examples of shock-acceleration of electrons that span a Mach number range that enters the high-Mach number regime of young SNR shocks, which does not occur at other heliospheric shocks frequently encountered by spacecraft (Sulaiman et al. 2015). These data represent an opportunity to determine the conditions under which electron acceleration occurs, over a range of Mach numbers.

An initial question posed by the presented results is: Why was a signature of electron acceleration only resolved at 31 of 508 Cassini crossings of Saturn's bow shock? Figure 1 suggests that electron acceleration signatures are more likely (but not exclusively) present when the shock was closer to the planet than is typical. The position of the shock is primarily controlled by the dynamic pressure (momentum flux) of the solar wind, $P_{SW}$, which is the product of the solar wind mass density and the square of the solar wind speed. These two upstream parameters are not continuously measured by Cassini due to instrument pointing constraints. However, Cassini studies to date have shown how (and to what extent) the influence of variations in key parameters can be assessed, even when dealing with hundreds of events (Masters et al. 2011; Sulaiman et al. 2015).

Figure 4a shows the square root of the "normalized" solar wind dynamic pressure on the *y*-axis against upstream magnetic field strength on the *x*-axis. Each solar wind dynamic pressure value was calculated by taking the crossing location and applying a semi-empirical model (Went et al. 2011), and then "normalized" to use the component of the upstream flow velocity normal to the shock surface (where the shock normal is also predicted by the model). Straight lines through this log-log parameter space describe loci of points at constant Alfvén Mach number ($M_A$), as shown by Sulaiman et al. (2015). Figure 4a shows the tendency for shock-acceleration of electrons under high dynamic pressure that we noted earlier, but does not indicate any clear dependence on $M_A$. Note that we cannot separate the dependence on upstream mass density from that on upstream flow speed.

At this stage we can appeal to current theories of the physics underlying electron "injection" at collisionless shocks in space plasmas. Different mechanisms have been proposed (e.g., Levinson 1992; Amano & Hoshino 2010; Riquelme & Spitkovsky 2011; Matsumoto et al. 2012, 2013; Kang et al. 2014; Guo & Giacalone 2015; Kato 2015; Matsukiyo & Matsumoto 2015). Common aspects of many of these proposed mechanisms are interactions between reflected thermal electrons and self-generated whistler waves just upstream of the shock. Oka et al. (2006) and Amano & Hoshino (2010) derived similar conditions for the resulting efficient "injection" of electrons, dependent on both the shock angle and Alfvén Mach number.

Figure 4b shows all the Cassini shock crossings in $\theta_{Bn}$-$M_A$ parameter space. The curved line gives the approximate electron injection threshold based on Oka et al. (2006; similar to the threshold presented by Amano & Hoshino (2010)). Below this curve the conditions at the shock are predicted to prohibit any efficient injection of electrons into the main acceleration process, whereas above the curve conditions are predicted to lead to efficient electron injection. Uncertainties on $\theta_{Bn}$ measurements are of order 10°, and uncertainties associated with $M_A$ are typically 25% (Masters et al. 2011). The 31 Cassini shock crossings where there is evidence of high-energy electrons all lie within the "injection-allowed" region to within errors, consistent with the underlying theory that predicts no electron injection problem at high-Mach number shocks (i.e., injection at any $\theta_{Bn}$).

However, a key question remains concerning this interpretation: Why are there Cassini crossings of Saturn's bow shock with no associated signature of electron injection that lie in the region of parameter space where such injection seems to be allowed? An explanation may be provided by the downstream thermal electron distributions measured by ELS. Figure 5a shows ELS spectra immediately downstream of the shock front for all acceleration events, as well as at all crossings without identified acceleration for which we have the highest confidence that they correspond to conditions are in fact predicted to allow efficient injection (gray-shaded region of Figure 4b). Figure 5b shows the average ELS spectrum of cases with and without evidence for electron injection in the LEMMS data.

The intensity peak/inflection at an energy of order 100 eV in the ELS spectrum for each crossing shown in Figure 5a corresponds to the ambient thermal electron population. The value of this "thermal energy" and the corresponding intensity shows significant differences between crossings, since these properties of the spectrum are controlled by the highly variable upstream solar wind conditions. Above this thermal energy the spectrum extends smoothly to higher energies where the intensities are lower. This higher energy part of the spectrum is generally well-captured by a power law, consistent with a non-thermal injected electron population (e.g., Oka et al. 2006).

This interpretation of the ELS data provides a potential explanation, which is illustrated in Figure 5b. The average ELS spectrum of crossings without an associated LEMMS signature (the gray curve) is described by a power law at energies above the thermal energy that has a similar slope to the average spectrum of crossings that were associated with a signature in LEMMS. However, extending these power laws to energies above 18 keV (i.e., into the higher LEMMS energy range) implies intensities at such energies that are below LEMMS channel background levels in the case of no LEMMS signature, in contrast with the cases that are associated with a LEMMS signature (see also Figure 3). Therefore, electron injection may have been taking place at all these crossings where such injection is predicted by theory, but the injected population may have been below the LEMMS background due to unfavorable prevailing upstream solar wind conditions at the crossing time.

5. Summary

Cassini spacecraft observations of electron acceleration at Saturn's bow shock are consistent with theory of electron injection at collisionless shocks most recently discussed by Amano & Hoshino (2010), which involves resonant interactions between thermal electrons and self-generated whistler waves just upstream of the shock. The broader implication of this is that the "pre-acceleration" of sub-relativistic electrons to higher (mildly-relativistic) energies at which they may undergo further acceleration via DSA is independent of shock angle at very high Alfvén Mach numbers, similar to those of young SNR shocks. This study has highlighted three Cassini shock crossings where a particularly strong signature of electron acceleration was measured by LEMMS. These events will be the subject of a dedicated future study that compares electron acceleration at quasi-parallel and quasi-perpendicular shocks, with an emphasis on the higher energy (DSA-like) acceleration process that produces relativistic particles.

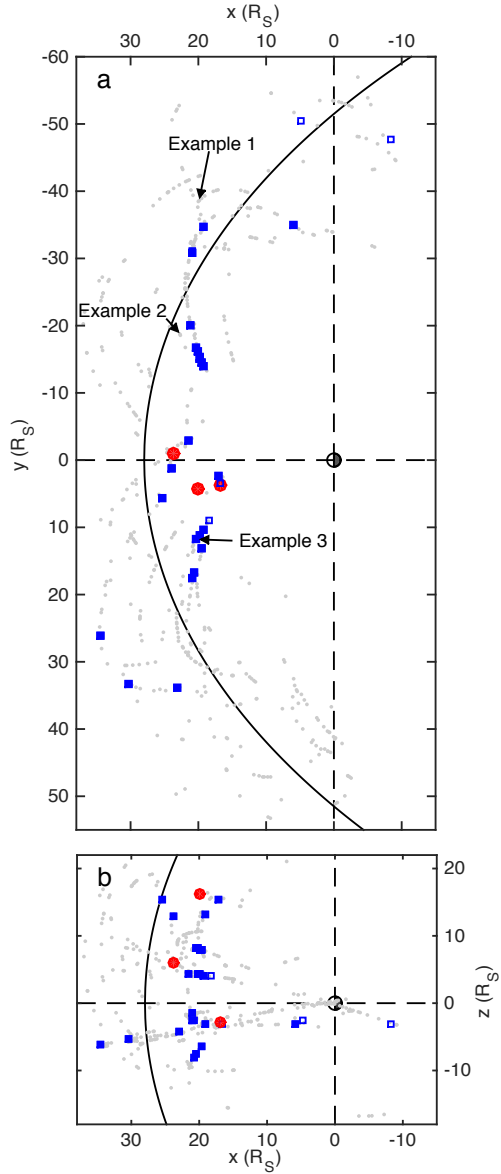

Figure 1. Locations of Cassini crossings of Saturn's bow shock included in the electron acceleration survey. Coordinate system: Origin at the center of Saturn, *x*-axis points toward the Sun, northward-directed *z*-axis defines an *xz* plane that contains the planet's magnetic dipole axis, *y*-axis completes the right-handed orthogonal set. Units: Saturn radii ($R_S$; 1 $R_S$ = 60268 km). (a) Crossing locations in the *xy* plane. (b) Crossing locations in the *xz* plane. Gray dots, blue squares, and red circles represent crossings with no LEMMS electron signature, a weak signature, and a strong signature, respectively (see Section 2). Unfilled blue squares are cases where misinterpretation of leaked magnetospheric electrons is plausible (see Section 2). The solid black curve in both panels gives the mean location of the shock surface (Went et al. 2011).

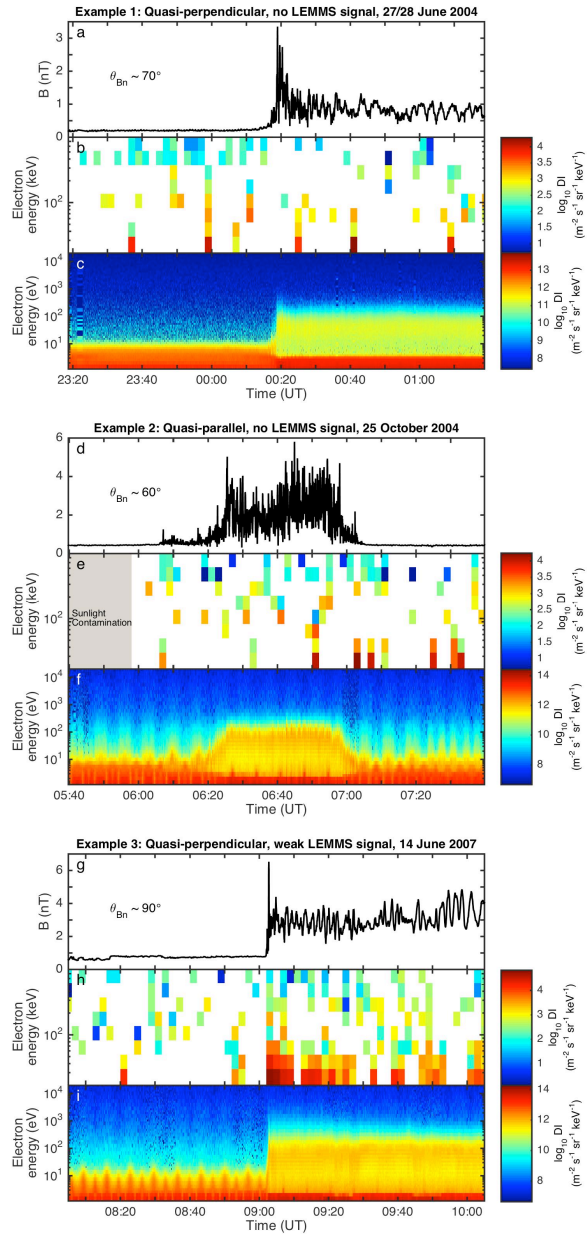

Figure 2. Data taken by Cassini during example crossings of Saturn's bow shock. (a-c) Typical quasi-perpendicular shock crossing made on 27 June 2004 with no LEMMS electron signal. (d-f) Shock crossing with lower shock angle ($\theta_{Bn}$) made on 25 October 2004, also with no LEMMS signal. (g-i) Quasi-perpendicular shock crossing made on 14 June 2007 with an associated LEMMS signature. MAG data are shown in panels a, d, and g. Background-subtracted LEMMS electron data are shown in panels (b, e, and h). ELS data (without background subtraction) are shown in panels c, f, and i, where modulation at ~5-minute period is due to sensor actuation. All intensities are given in Differential Intensity (DI).

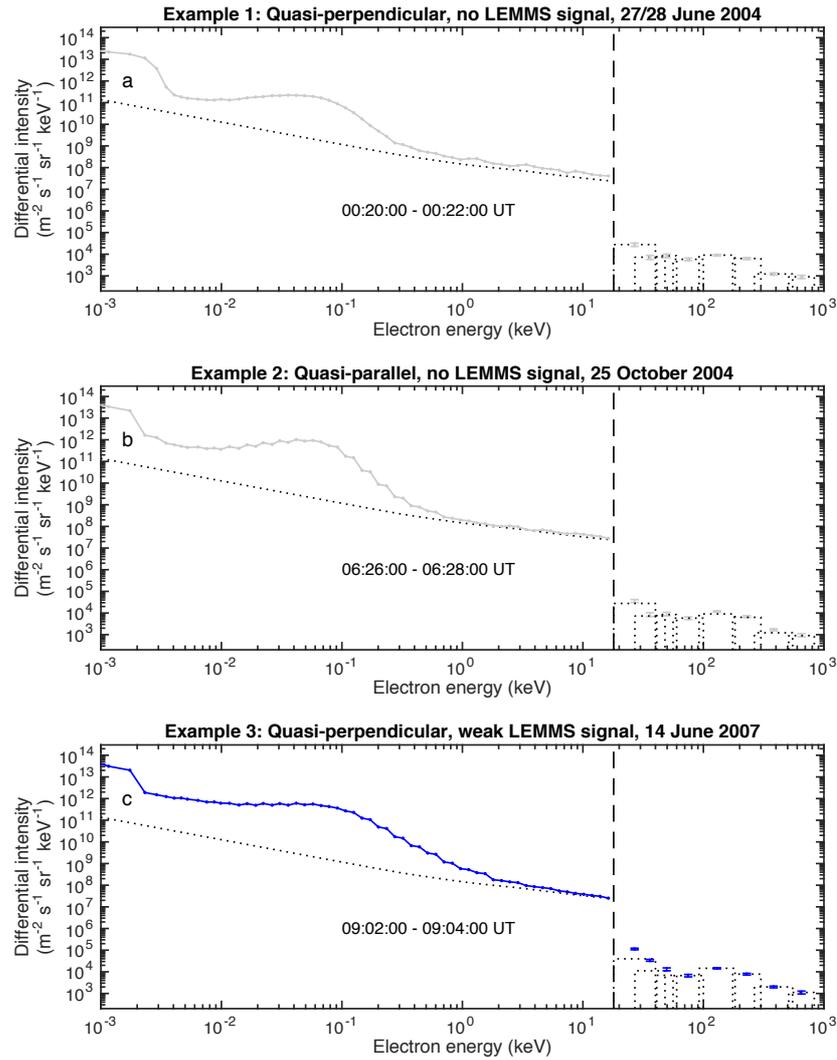

Figure 3. Combined ELS-LEMMS two-minute-averaged electron spectra for the three example shock crossings shown in Figure 2. (a) Example 1, quasi-perpendicular, no LEMMS electron signal. (b) Example 2, atypical magnetic signature, no LEMMS signal. (c) Example 3, quasi-perpendicular, LEMMS signal. ELS energy range upper limit set as 18 keV (lower limit of lowest LEMMS energy channel). "Step-like" features in ELS spectra are due to onboard spacecraft averaging in response to telemetry constraints. The dotted curve below 18 keV is the ELS background, whereas the dotted rectangles above 18 keV give the both the energy range and background level of each LEMMS electron channel. The intensity of each LEMMS electron channel is given by a data point with vertical error bars, located at an energy in the middle of the channel energy range (using a logarithmic scale).

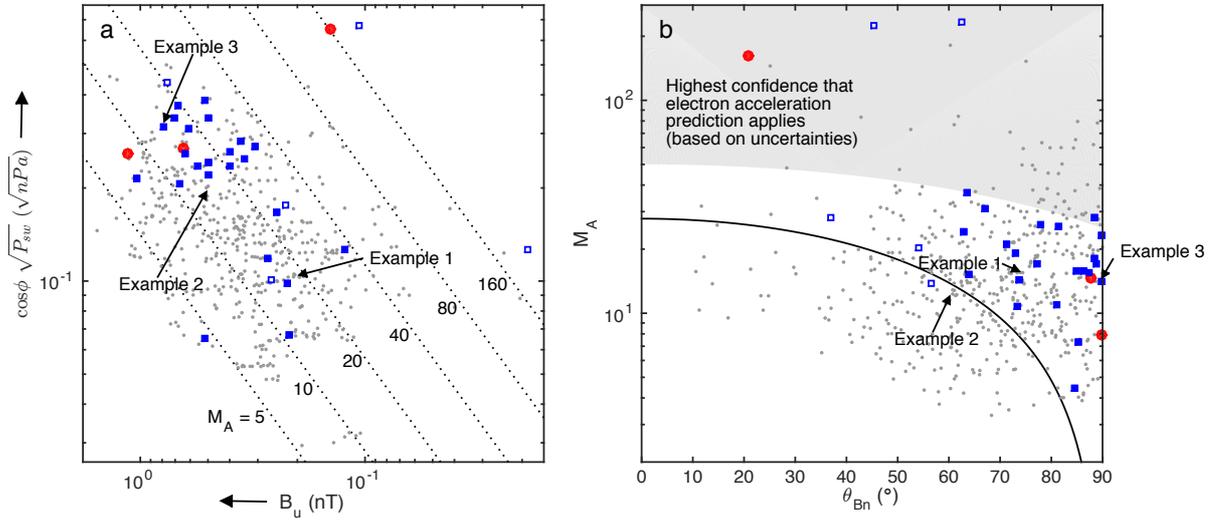

Figure 4. Assessing the parameter dependence of electron acceleration at Saturn's bow shock. (a) Shock crossings organized by "normalized" solar wind dynamic pressure ($P_{sw}$) and upstream magnetic field strength ($B_u$) (see Section 4). (b) Shock crossings organized by Alfvén Mach number ($M_A$) and shock angle ($\theta_{Bn}$) (see Section 4), where the black curve denotes the efficient electron injection threshold following Oka et al. (2006) (injection predicted above the curve, whereas not predicted below). Gray dots, blue squares, and red circles represent crossings with no LEMMS electron signature, a weak signature, and a strong signature, respectively (see Section 2). Unfilled blue squares are cases where misinterpretation of leaked magnetospheric electrons is plausible (see Section 2).

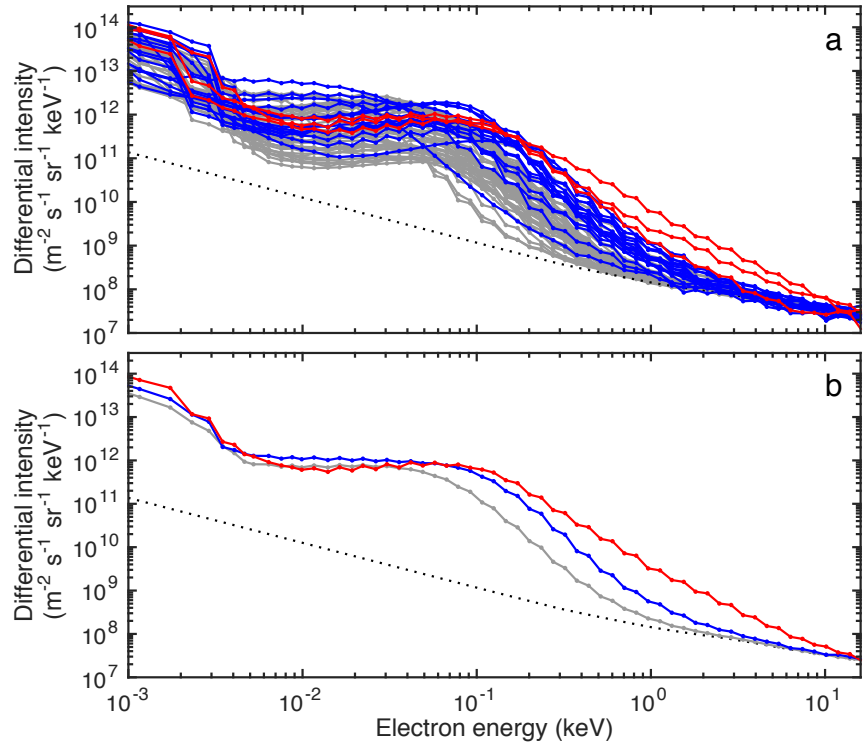

Figure 5. ELS spectra measured immediately downstream of Saturn's bow shock under conditions where electron acceleration is predicted (but not necessarily observed, see Section 4). (a) All spectra. (b) Energy-averaged spectra separated by category of LEMMS signature: Gray, blue, and red represent crossings with no LEMMS electron signature, a weak signature, and a strong signature, respectively (see Section 2). The dotted curve is the ELS background.

Acknowledgements

We thank Cassini instrument Principal Investigators S. M. Krimigis, D. T. Young, and J. H. Waite. This work was supported by UK STFC through rolling grants to MSSL/UCL and Imperial College London. LS was supported by Polish NSC grant DEC-2012/04/A/ST9/00083. AHS is supported by NASA through Contract 1415150 with Jet Propulsion Laboratory.

References

- Abdo, A. A., Ackermann, M., Ajello, M., et al. 2011, ApJ, 734, 9

- Achilleos , N., Bertucci, C., Russell, C. T., et al. 2006, JGRA, 111, A03201

- Ackermann, M., Ajello, M., Atwood, W. B., et al. 2010, PhRvD, 82, 092004

- Aharonian, F. A., Akhperjanian, A. G., Aye, K.-M., et al. 2004, Natur, 432, 75

- Amano, T. & Hoshino, M. 2010, PhRvL, 104, 181102

- Bell, A. R. 2013, APh, 43, 56

- Blandford, R. & Eichler, D. 1987, PhR, 154, 1

- Burgess, D. 2007, LNP, 725, 161

- Dougherty, M. K., Kellock, S., Southwood, D. J., et al. 2004, SSRv, 114, 331

- Drury, L. Oc. 1983, RPPh, 46, 973

- Gosling, J. T., Thomsen, M. F., Bame, S. J., et al. 1989, JGR, 94, 10011

- Guo, F., & Giacalone, J. 2015, ApJ, 802, 8

- Helder, E. A., Vink, J., Bykov, A. M., et al. 2012 SSRv, 173, 369.

- Horbury, T. S., Cargill, P. J., Lucek, E. A. 2002 JGRA, 107, A8.

- Jones, F. C., & Ellison, D. C. 1991, SSRv, 58 259

- Kanani et al. 2010, JGRA, 118, 1620

- Kang, H., Petrosian V., Ryu D. Jones, T. W. 2014, ApJ, 788, 142

- Kato, T. N. 2015, ApJ, 802, 115

- Krimigis, S. M. 1992, SSRv, 59 167

- Krimigis, S. M., Mitchell, D. G., Hamilton, D. C., et al. 2004, SSRv, 114, 233

- Levinson, A. 1992, ApJ, 401, 73

- Masters, A., Schwartz, S. J., Henley, E. M., et al. 2011, JGRA, 116, A10107

- Masters, A., Stawarz, L., Fujimoto, M., et al. 2013, NatPh, 9, 164

- Matsukiyo, S., & Matsumoto, Y. 2015, JPhCS, 642, 012017

- Matsumoto, Y., Amano, T., Hoshino, M. 2012, ApJ, 755, 11

- Matsumoto, Y., Amano, T., Hoshino, M. 2013, PhRvL, 111, 215003

- Oka, M., Terasawa, T., Seki, Y., et al. 2006, GeoRL, 33, L24104

- Reynolds, S. P. 2008, ARA&A, 46, 89

- Riquelme, M. A., & Spitkovsky, A. 2011, Ap J, 733, 15

- Russell, C. T. 1985, in Collisionless Shocks in the Heliosphere: Reviews of Current Research, ed. Tsurutani, B. T., & Stone, R. G. (Washington DC American Geophysical Union), 109

- Sarris, E. T., & Krimigis, S. M. 1985, ApJ. 298 676

- Sergis, N., Jackman, C. M., Masters, A., et al. 2013, JGRA, 118, 1620

- Shimada, N., Terasawa, T., Hoshino, M., et al. 1999, Ap&SS, 264, 481

- Smith, E. J. 1985, in Collisionless Shocks in the Heliosphere: Reviews of Current Research, ed. Tsurutani, B. T., & Stone, R. G. (Washington DC American Geophysical Union), 69

- Sulaiman, A. H., Masters, A., Dougherty, M. K., et al. 2015, PhRvL, 115, 125001

- Treumann, R. A. 2009, A&ARv, 17, 409

- Uchiyama, Y., Aharonian, F. A., Tanaka, T., et al. 2007, Natur, 449, 576

- Went, D. R., Hospodarsky, G. B., Masters, A., et al. 2011, JGRA, 116, A07202

- Young, D. T., Berthelier, J. J., Blanc, M., et al. 2004, SSRv, 114, 1